\documentclass[twocolumn,showpacs,preprintnumbers,amsmath,amssymb,prb]{revtex4} 
\usepackage{tabularx} 
\usepackage{graphicx} 
\usepackage{longtable} 
\usepackage{dcolumn} 
\usepackage{setspace}
\usepackage{bm}
\usepackage{wrapfig}                                                  
\usepackage{color}
\usepackage{fancyhdr} 
\usepackage{multirow}

\definecolor{Remarks}{rgb}{1,0.3,0.3}

\newcommand\COMMENTED[1] {}

\newcommand\GAUSSIAN[1][]{{\footnotesize{GAUSSIAN#1}}}

\newcommand\QUANTESPRESSO     {{\footnotesize{QuantumESPRESSO}}}

\newcommand\Ref[1]     {Ref.~\onlinecite{#1}}

\begin{document}

\title{ First-principles calculations of $^{17}$O NMR chemical
  shielding\\in Pb(Zr$_{1/2}$Ti$_{1/2}$)O$_3$ and
  Pb(Mg$_{1/3}$Nb$_{2/3}$)O$_3$: linear dependence on
  transition-metal/oxygen bond lengths }

\author{Daniel L. Pechkis, Eric J. Walter and Henry Krakauer}
\affiliation{Department of Physics, College of William and Mary,
  Williamsburg, VA 23187-8795.}

\date{\today}

\begin{abstract}

First-principles density functional theory (DFT) oxygen chemical shift
tensors were calculated for A(B,B$^{\prime}$)O$_3$ perovskite alloys
Pb(Zr$_{1/2}$Ti$_{1/2}$)O$_3$ (PZT) and Pb(Mg$_{1/3}$Nb$_{2/3}$)O$_3$
(PMN).  Quantum chemistry methods for embedded clusters and the GIPAW
method~[C.\ J.\ Pickard and F.\ Mauri, {\it Phys. Rev. B} {\bf 63}
  245101 (2001)] for periodic boundary conditions were used.  Results
from both methods are in good agreement for PZT and prototypical
perovskites. PMN results were obtained using only GIPAW.  Both
isotropic $\delta_\mathrm{iso}$ and axial $\delta_\mathrm{ax}$
chemical shifts were found to vary approximately linearly as a
function of the nearest-distance transition-metal/oxygen bond length,
$r_\mathrm{s}$.  Using these results, we argue against Ti clustering
in PZT, as conjectured from recent $^{17}$O NMR magic-angle-spinning
measurements.  Our findings indicate that $^{17}$O NMR measurements,
coupled with first-principles calculations, can be an important probe
of local structure in complex perovskite solid solutions.
\end{abstract}

 
 
\maketitle

\section{Introduction} 
\linespread{1.0}

High performance solid solution ferroelectrics, based on the ideal
ABO$_3$ perovskite structure, are widely used in technological
applications such as ultrasonic transducers, sensors, actuators, and
thin film applications.~\cite{2005-RMP-FE-thinfilms,Ref:Scott_appl}
The strong electromechanical coupling in these materials is related to
a balance of competing instabilities, such as cation off-centerings
and oxygen octahedral rotations.  The perovskite structure offers many
ways to fine tune these interactions through chemical substitutions,
such as alloying on the A and/or B sites, and through epitaxial
control in layered and thin film
geometries.~\cite{2005-RMP-FE-thinfilms}

Solid state nuclear magnetic resonance (NMR) has increasingly been
used to study the local structure and dynamics of these complex
perovskites.~\cite{ref:ZhouHoatsonVold-Pb,ref:Baldwin05,ref:VijayakumarHoatsonVold-PMN,ref:BT-ST-O_Blinc}
NMR spectra of a target nucleus are largely determined by the coupling
of its magnetic dipole and electric quadrupole moments with the local
magnetic field and electric field gradient, respectively. The
interpretation of NMR spectra in complex solid solution perovskites is
complicated by the presence of broad spectral features due to
disorder.  First-principles calculations of electric field
gradient~\cite{ref:mao014105} and chemical
shielding~\cite{ref:pechkis1} tensors can play an important role
interpreting NMR spectra in these materials.

Previously we demonstrated a near linear dependence of the chemical
shielding tensor $\hat{\sigma}$ on the oxygen nearest neighbor B-O
bond distance $r_\mathrm{s}$ for the prototypical perovskites
BaTiO$_3$ (BT), SrTiO$_3$ (ST), PbTiO$_3$ (PT), and PbZrO$_3$
(PZ).~\cite{ref:pechkis1} The linear dependence was shown to arise
from large paramagnetic contributions to $\sigma_{x}$ and $\sigma_{y}$
principal values (our convention identifies the ``$z$" principal axis
as that most nearly parallel to the B-O-B bond direction), due to
virtual transitions between O(2p) and unoccupied B($n$d) states.  This
linear variation is confirmed here for two complex perovskite solid
solutions, PZT and PMN.

First-principles chemical shielding calculations have traditionally
been done with the embedded cluster approach, using standard quantum
chemistry methods~\cite{Gaussian98Abbrev,Gaussian03Abbrev,Gaussian09Abbrev} with
gaussian type orbitals (GTO).  More recently, the planewave based
GIPAW method with PBC has provided an alternative approach.
Relatively few calculations for transition metal oxides have been
reported using either technique.  Here we use complementary
calculations with both methods to cross validate convergence with
respect to cluster size and termination effects, basis sets, and the
accuracy of pseudopotentials (PSPs).

Quantum-chemistry methods can calculate chemical shielding tensors for
embedded clusters, using a range of approximations, from Hartree Fock
and density functional theory (DFT) with semilocal or hybrid
exchange-correlation functionals, to explicitly correlated methods
such Moller-Plesset perturbation theory and coupled cluster
approaches. \cite{1999-Helgaker,2004-bookkaupp,2007-Vaara} The
principal difficulties with the embedded cluster approach are
controlling size and basis set convergence.  Size effects can be
monitored by studying increasingly larger clusters. Long-range
electrostatic interactions can be handled by embedding the cluster in
large arrays of point charges and eliminating depolarizing fields, as
described in \Ref{ref:pechkis1}.  Achieving the basis set limit can be
problematic in some cases, because atom-centered GTO's do not form a
complete orthogonal basis.  Nevertheless, basis set convergence is
generally well controlled through the use of standardized GTO basis
sets. \cite{EMSL_BasisSets2007} The cluster approach becomes
inefficient, however, for complex systems, because separate cluster
calculations are usually required for each inequivalent target atom.

GIPAW calculations are naturally adapted to ordered crystalline
solids, since size convergence is effectively achieved by using
primitive unit cells with well-converged $k$-point quadrature grids
for Brillouin zone integrations.  Disordered solids can be treated
using supercells.  Planewaves form a complete basis, so convergence to
the basis set limit is straightforward.  The method applies the
projector-augmented-wave (PAW) reconstruction~\cite{ref:PAWrecon} to a
conventional PSP calculation to obtain all-electron valence wave
functions, which are required for accurate calculations of the
chemical shielding.  There are two principle issues with the GIPAW
method.  GIPAW PSPs are more difficult to construct than standard
norm-conserving PSPs.  To achieve good transferability, they may
require multiple partial wave channels and large planewave cutoff
energies for some target atoms.  The construction of the PAW
atomic-like augmentation basis also requires care.  To date, GIPAW
calculations have been carried out for only a limited number of
transition metal oxide systems.\cite{ref:V-NMR-CASTEP,
  ref:Middlemiss2010}

PZT is a homovalent mixture of Ti$^{4+}$ and Zr$^{4+}$ transition
metal cations, while PMN is a heterovalent 2:1 mixture of the
Nb$^{5+}$ transition metal cation and the Mg$^{2+}$ alkaline earth
cation.  Both embedded cluster and GIPAW calculations were carried out
for PZT, while PMN results were obtained only with the GIPAW method,
using PBC.  First-principles relaxed structural models were used to
simulate PZT and PMN structures.
 
The remainder of the paper is organized as follows. The theoretical
approaches are described in Sec.~\ref{sec:method}.  Results and
discussion are presented in Secs.~\ref{Results} and~\ref{Discussion},
respectively.  We summarize and conclude in Sec.~\ref{summary}.
  
\begin{table*}[t]
\begin{center}
\caption{Pseudopotential construction parameters, see text for description.}  
\begin{tabular*}{0.98\textwidth}{@{\extracolsep{\fill}}llclc}\hline \hline 
  & \multicolumn{1}{c}{reference state}  & r$_c$(au) &    \multicolumn{1}{c}{reference energies (Ry)} & core correction radius (au)\\\hline
O & {\bf 2s$^2$},2p$^6$,{\it 3d,3s,3p} & 1.2,1.2,1.5 & *,*,0.10,0.10,0.10   & 0.34  \\
Ti & {\bf 3s$^2$},3p$^6$,3d$^0$,{\it 4s,4p,4d} & 0.9,0.9,0.9 & *,*,*,-2.73,0.50,0.10 & 0.53 \\
Pb & {\bf 6s$^2$},6p$^0$,5d$^{10}$,{\it 7s,7p,6d} & 2.0,2.4,1.0 & *,*,*,-1.00,-0.20,-1.80 & 0.89  \\
Zr & {\bf 4s$^2$},4p$^6$,4d$^0$,{\it 5s,5p,5d} & 1.0,1.2,1.4 & *,*,*,-2.45,-2.00,-1.37 & 0.80 \\
Sr & 4s$^2$,{\bf 4p$^6$},4d$^{0}$,{\it 5s,5p} & 1.2,1.3,1.7 & *,*,-1.00,-0.72 & 0.88 \\
Ba & {\bf 5s$^2$},5p$^6$,{\it 5d,6s,6p} & 1.5,1.7,2.0 & *,*,-0.95,-0.90,-1.50 & 1.19 \\
Mg & {\bf 2s$^2$},2p$^6$,{\it 3d,3s,3p} & 0.6,0.6,1.5 & *,*,0.3,-1.3,0.1 & 0.30 \\ 
Nb & {\bf 4s$^2$},4p$^6$,4d$^0$,{\it 5s,5p,5d} & 1.0,1.1,1.3 & *,*,*,-3.30,-2.70,-1.27 & 0.80 \\ 
K & {\bf 3s$^2$},3p$^6$,3d$^0$,{\it 4s,4p} & 1.4,1.5,1.5 & *,*,*,-1.50,-0.80 & 0.50 \\ \hline
\end{tabular*}
\vspace{-10pt}
\label{psp}
\end{center}
\end{table*}

\section{Theoretical Methods}
\label{sec:method} 

The chemical shielding tensor $\hat{\sigma}$ determines the total
magnetic field at an atomic nucleus,
\begin{equation}
\mathbf{B} =
(1-\hat{\sigma})\mathbf{B}_{\mathrm{ext}}
\, ,
\label{eq:B_ind}
\end{equation}
where $\mathbf{B}_{\mathrm{ext}}$ is the external field.  For the
systems considered here, $\hat{\sigma}$ is calculated using embedded
cluster and GIPAW-PBC methods. The symmetric
\cite{Asymm-sigma1968,Asymm-sigma1991} $\hat{\sigma}$ tensor is
determined by its principal axis components, with isotropic and
anisotropic parts, conventionally defined as \cite{baugher1969}
\begin{equation}
\begin{array}{c}
 \sigma _{{\rm{iso}}}  = {1 \over 3} \left( {\sigma _x  + \sigma _y  + \sigma _z } \right)  
                       = {1 \over 3} \rm{Tr}\,\hat{\sigma}\\ 
 \sigma _{{\rm{ax}}}  =  {1 \over 6} \left( {2\sigma _z  - \sigma _x  - \sigma _y } \right)
                       = {1 \over 2} \left( {\sigma _z - \sigma _{{\rm{iso}}} } \right)   \\ 
 \sigma _{{\rm{aniso}}}= {1 \over 2}\left( {\sigma _y  - \sigma_x        } \right) \\ 
 \end{array}
\label{eq:sigmaiso}
\end{equation}
As mentioned, our convention for the perovskite structure identifies
the ``$z$'' principal axis as that most nearly parallel to the
\mbox{B-O-B} bond direction.  NMR measurements of $\hat{\sigma}$ are
usually reported with respect to a reference material, where the
chemical shift tensor $\hat{\delta}$ is defined
as~\cite{2004-bookkaupp}
\begin{equation}
\hat{\delta} =  -(\hat{\sigma}  - \sigma_\mathrm{ref}),
\label{eq:delta}
\end{equation}
with corresponding definitions to those in Eq.~\ref{eq:sigmaiso}.
The experimental reference for oxygen is liquid H$_2$O.  

The theoretical oxygen reference value $\sigma_{\mathrm{ref}}^{\mathrm
  {th.}}$ is determined from a linear regression of $\delta_{\rm
  iso}^{\rm expt.}$ versus $\sigma_{\rm iso}^{\rm theory}$.
\cite{ref:Charpentier,ref:Middlemiss2010} This yields the relation
\begin{equation}
  \delta^{\mathrm {th.}}                                  =  -m  (\sigma^{\mathrm {th.}} - \sigma_{\mathrm{ref}}^{\mathrm {th.}}).
\label{eq:delta_theory}
\end{equation}
Rather than regarding the slope $m$ as an independent fitting
parameter, we set \mbox{$m=1$}, which leaves
$\sigma_{\mathrm{ref}}^{\mathrm {th.}}$ as the only independent
parameter. Allowing $m$ to vary yields fits of similar overall
quality, but somewhat distorts chemical shift differences between
inequivalent oxygen sites within the same material. Constraining
\mbox{$m=1$} allows better cancellation of errors, due to systematic
effects such as the choice of methodological approach or DFT
exchange-correlation functional.

\subsection{Embedded cluster calculations}
\label{sec:cluster}
 
A detailed discussion of this approach is given in
Ref.\ \onlinecite{ref:pechkis1}. We briefly summarize some of the key
features of this method.  With central O atoms, embedded clusters
consisted of either 21``quantum'' (QM) atoms,
(A$_4$B$_2$O$_{15}$)$^{14-}$, or 65 QM atoms,
(A$_4$B$_{10}$O$_{51}$)$^{51-}$. In these clusters, all cation atoms
are fully coordinated with QM O atoms.  The QM clusters are further
surrounded by a large array of point charges, which reproduce the
Madelung potential in the QM region.~\cite{ref:ewald} To alleviate
artificial polarization of boundary O(2p) states, the nearest-neighbor
(nn) and the next-nearest-neighbor cation point charges of boundary O
atoms are replaced by ``empty'' PSPs
(ePSPs).~\cite{ref:pechkis1,ref:TIPs} Finally, in non-centrosymmetric
clusters, depolarizing electric fields are removed by applying an
external electric field. \cite{ref:pechkis1}

Calculations were performed with the {\GAUSSIAN} computational
package, \cite{Gaussian98Abbrev,Gaussian09Abbrev} and the chemical
shielding tensor was determined using the continuous set of gauge
transformations (CSGT) method. \cite{ref:CSGT,cheeseman} Calculations
were done using the DFT hybrid B3LYP~\cite{B3LYP}, as well as
generalized gradient approximations (GGA), using the
PW91~\cite{ref:PW91} and PBE~\cite{ref:pbe} forms.  Douglas-Kroll-Hess
2nd order scalar-relativistic calculations were performed on selected
systems.  Atom-centered GTO basis functions were associated with all
the QM atoms. All-electron treatments were used for the O and Ti
atoms, while the other QM atoms were represented using
scalar-relativistic small core (scalar-RSC) PSPs [also called
  effective core potentials (ECPs)].  The well-converged GTO basis
sets and ECPs used for these calculations are described in
\Ref{ref:pechkis1} and were taken from the EMSL
website.~\cite{EMSL_BasisSets2007} 

\subsection{GIPAW calculations}

Calculations with PBC used the GIPAW functionality of the
\QUANTESPRESSO~(QE) code.\cite{ref:QE} These calculations are done in
two steps.  A standard ground state norm-conserving PSP calculation is
first performed.  This is followed by a linear response calculation in
the presence of an external magnetic field.  The linear response
calculation uses all-electron like valence wave functions, which are
represented by planewaves, modified near the nuclei by atomic-like PAW
augmentation basis functions.  The PAW basis functions are constructed
when the PSP is generated, as further discussed below.  Calculations
reported below used well converged Monkhorst-Pack\cite{ref:MP}
Brillouin zone $k$-point sampling, {\em e.g.}
\mbox{6$\times$6$\times$6} for the prototypical perovskites ST, BT,
and PT.  Unless otherwise specified, the PBE GGA functional was used
in all QE calculations.

All PSPs were constructed using the ``ld1'' PSP generation code
(distributed with QE). All norm-conserving PSPs were
scalar-relativistic Troullier-Martins\cite{ref:TM}
type. Table~\ref{psp} shows the construction parameters used for all
PSPs employed in this paper. The local channel is indicated by
boldface type. States that are in italics were generated using a
Hamann type\cite{ref:Hamann} reference state. The $r_c$'s correspond
to the $s$, $p$, and $d$ channels respectively. The next column shows
the PSP reference energies for the $s$, $p$, and $d$ channels.  The
symbol `*' indicates that the corresponding all-electron eigenvalue
was used for this state.  It should be noted that these parameters are
for the PSP projectors used in the ground state self-consistent
total-energy calculations.  Parallel to the PSP construction, a second
atomic calculation is performed to generate the required GIPAW
augmentation basis functions, consisting of all-electron and pseudo
partial wave radial functions.  The number of GIPAW angular momentum
channels was the same as for the PSPs.  For all except Pb, O and K,
the augmentation basis functions were generated using the
corresponding all-electron eigenvalues ({\it n.b.}, the Ti$^{4+}$ 4$d$
state is bound, for example), rather than the values in
Table~\ref{psp}.  For Pb, O and K, the values in Table~\ref{psp} were
used, except for the Pb 7$s$, which used the all-electron eigenvalue.
The final column shows the non-linear core-correction radius for each
potential.  A conservative 350 Ry energy cutoff was used. This could
have been reduced by using larger r$_c$'s for the metal PSPs.  The 350
Ry energy cutoff, while high, was easily tractable for all systems
studied in this paper.  This cutoff yields chemical shieldings to
within about 1 ppm, as indicated by test calculations with other
settings.

\begin{table}[ht]
\begin{center}
\caption{Comparison of chemical shielding results, using embedded
  clusters and GIPAW-PBC from {QE} and \Ref{ref:Middlemiss2010}. The
  principle values of the oxygen chemical shielding tensor are
  presented for the TiO$_2$ molecule and rutile, cubic ST, cubic BT,
  and tetragonal PT.  All values are from GGA calculations. Embedded
  cluster results are labeled as C-$n$, where $n$ is is the number of
  QM atoms in the cluster. The GIPAW-PBC and C-65 calculations were
  done with relativistic PBE, while the C-21 calculations were done
  with non-relativistic PW91.  }
\begin{tabular*}{0.48\textwidth}{@{\extracolsep{\fill}}lcccc} \hline \hline 
~~~ &\\
       &       $\sigma_x$       &       $\sigma_y$       &       $\sigma_z$       &       $\sigma_{iso}$\\
\hline
\multicolumn{5}{l}{TiO$_2$ (molecule)}\\[1pt]
Gaussian  &       -1803       &       -801       &       148       &       -819       \\
GIPAW       &       -1826       &       -811       &       146       &       -830       \\ 
\hline
\multicolumn{5}{l}{TiO$_2$ (rutile)}              \\[1pt]
C-99       &       -499       &       -380       &       -306       &       -395       \\
GIPAW       &       -483       &       -380       &       -296       &       -386       \\
\hline
\multicolumn{5}{l}{SrTiO$_3$ (cubic)} \\       [1pt]                                                             
C-21       &       -353       &       -353       &       46       &       -220       \\ 
C-65       &       -403       &       -403       &       27       &       -260       \\
GIPAW        &       -429       &       -429       &       7       &       -284 / -287\footnotemark[1]      \\
\hline
\multicolumn{5}{l}{BaTiO$_3$ (cubic)} \\       [1pt]                                                             
C-21       &       -414       &       -414       &       49       &       -260       \\
C-65       &       -483       &       -483       &       22       &       -315       \\
GIPAW        &       -529       &       -529       &       -31       &       -363 / -379\footnotemark[2]        \\
\hline
\multicolumn{5}{l}{PbTiO$_3$-axial O (P4mm)} \\[1pt]                                                                    
C-21       &       -562       &       -562       &       123       &       -334       \\
C-65       &       -599       &       -599       &       67       &       -377       \\
GIPAW       &       -630       &       -630       &       77       &       -394       \\ 
\hline
\multicolumn{5}{l}{PbTiO$_3$-equatorial O (P4mm)} \\[1pt]                                                                    
C-21       &       -286       &       -228       &       -32       &       -182       \\
C-65       &       -365       &       -277       &       -35       &       -226       \\
GIPAW       &       -398       &       -284       &       -23       &       -235       \\
\hline \hline
\end{tabular*}
\footnotetext[1]{Ref.\ \onlinecite{ref:Middlemiss2010} determined $\sigma_{iso}$ using the experimental structure}
\footnotetext[2]{Ref.\ \onlinecite{ref:Middlemiss2010} determined $\sigma_{iso}$ using the relaxed structure}
\label{tab:prototypical}
\end{center}
\end{table}

\begin{table*}[ht]
\begin{center}
\caption{
The derived theoretical
oxygen reference value, $\sigma_{\mathrm{ref}}^{\mathrm{th.}}$, is
used to determine theoretical isotropic chemical shifts from the
corresponding calculated isotropic chemical shielding values. 
Experimental chemical shifts are shown for comparison.
The values of $\sigma_{\mathrm{ref}}^{\mathrm{th.}}$
are shown in the third row from the bottom of this table. The rms error and maximum
deviation of the calculated shifts, compared to experiment, are also shown. 
Calculated cluster results (21 and 65 QM atoms) with B3LYP and PW91
exchange-correlation are shown together with GIPAW-PBC with PBE
exchange-correlation. (PZ experimental chemical shift site assignments were corrected in \Ref{ref:pechkis1} and are used here.)
}
\begin{tabular*}{0.98\textwidth}{@{\extracolsep{\fill}}llcccccccc} \hline \hline 
& & Expt\footnotemark[1] & \multicolumn{4}{c}{Cluster}&\multicolumn{3}{c}{PBC-GIPAW} \\ 
 & & & \multicolumn{2}{c}{B3LYP}&\multicolumn{2}{c}{PW91}&present \footnotemark[6]&present&other \\ 
& & & 21 & 65 & 21 & 65 &  &  &  \\ \hline
& & & & & & & & & \\ 
ST (cubic)& & $467 \pm 5$ & 491 & 477 & 494 & 491 & 491 & 496 & 503 \footnotemark[2]	\\ \hline
& & & & & & & & & \\ 
BT (cubic)& & $546 \pm 5$ & 536 & 537 & 534 & 547 & 570 & 575 &	595 \footnotemark[3]	\\ \hline
& & & & & & & & & \\ 
BT (P4mm) & & & & & & & & & \\
          & O$_{\rm ax}$	& $570 \pm 5$ & 591 & &	571 & & & 579 & 573 \footnotemark[4] / 611 \footnotemark[5] \\
          & O$_{\rm eq}$ & $520 \pm 5$ & 515 & & 516 & &	& 535 & 563 \footnotemark[4] / 531 \footnotemark[5] \\
          & $\Delta$(O$_{\rm ax}$ -- O$_{\rm eq}$) & 50 & 76 & & 56 & & &	44 & 10 / 80 \\ \hline
& & & & & & & & & \\ 
PT (P4mm) & & & & & & & & & \\
          & O$_{\rm ax}$	& $647 \pm 2$ & 644 & 644 & 608 & 609 & 601 & 606 &   \\ 
          & O$_{\rm eq}$	& $443 \pm 2$ & 449 & 445 & 457 & 455 & 442 & 447 & \\ 
          & $\Delta$(O$_{\rm ax}$ -- O$_{\rm eq}$)  & 204 & 195 & 199 & 152 & 154 & 159 &	159 & \\ \hline
& & & & & & & & & \\ 
PZ (Pbam) & & & & & & & & & \\
          & O1-4g & $365 \pm 2$ & 357 & & 367 & & & 355 & \\ 
          & O1$^{\prime}$-4g & $351 \pm 2$ & 346 & & 356 & & & 336 & \\
          & O2-8i & $356 \pm 2$ & 355 & & 364 & & & 349 & \\ 
          & O3-4f & $329 \pm 2$ & 324 & & 336 & & & 309 & \\ 
          & O4-4e & $408 \pm 2$	& 392 & & 399 & & & 415 & \\ 
          &$\Delta$(O4 -- O1)& 43 & 35 & & 31 & & & 60 &\\ 
          &$\Delta$(O4  -- O1$^{\prime}$)& 57 & 47 & & 43 & & & 79 & \\ 
          & $\Delta$O4 -- O2)& 52 & 38 & & 35 & & & 66 &\\ 
          &$\Delta$(O4 -- O3)& 79 & 68 & & 63 & & & 105 &\\
& & & & & & & & & \\ \hline
  $\sigma_{\mathrm{ref}}^{\mathrm {th.}}$          &  & 288 \footnotemark[5] & 293 & 238 & 275 & 217 & 207 & 212 & 216 \\ 
RMS error & & & 12 & 7 & 16 & 23 & 28 & 20 & 26 \\ 
Max Dev \footnotemark[6]& & & 24 & 10 & 39 & 38 & 46 & 41 & \\ \hline
\hline 
\end{tabular*}
\begin{minipage}[b]{\textwidth}
\footnotetext[1]{ST and BT experimental chemical shift values are from
  Ref.\ \onlinecite{ref:BT-ST-O_Blinc} and the PT and PZ experimental
  chemical shift values are from Ref.
  \onlinecite{ref:Baldwin05}. }
\footnotetext[2]{Ref.\ \onlinecite{ref:Middlemiss2010} determined $\sigma_{iso}$ using the experimental structure}
\footnotetext[3]{Ref.\ \onlinecite{ref:Middlemiss2010} determined $\sigma_{iso}$ using the relaxed structure}
\footnotetext[4]{The $^{17}$O experimental chemical shielding reference is  liquid water,  $\sigma_\mathrm{iso}^{\mathrm{water}}=287.5$~ppm~\cite{ref:wasylishen}}.
\footnotetext[5]{Max Dev = max(abs($\delta^{\rm theory} -  \delta^{\rm
    expt.}$)) }
\footnotetext[6] {Used $\sigma_{\mathrm{ref}}^{\mathrm {th.}}$
  determined from linear regression on the same systems as the 65
  QM atom cluster-PW91 results (see text).}
\label{tab:O-shifts_prototypical}
\end{minipage}
\end{center}
\end{table*}

\subsection{Comparison of methods}

Table \ref{tab:prototypical} shows the comparison between the GIPAW
and cluster approach for the TiO$_2$ molecule, the rutile crystalline
solid, as well as some results for prototypical perovskites. Where
available, published GIPAW results from \Ref{ref:Middlemiss2010} are
also given for comparison.

TiO$_2$ molecule calculations were for a Ti-O bond length of
1.651\,{\AA} and a O-Ti-O angle of 114.2$^{\circ}$. With PBC-GIPAW, a
\mbox{22$\times$24$\times$28} Bohr supercell was used. The Gaussian
calculation was for the isolated molecule and used a basis set of
cc-pwCVQZ-DK and IGLO-III for Ti and O, respectively.  The
experimental rutile structure was used for solid
TiO$_2$.\cite{ref:TiO2_rutile_lattice} The PBC-GIPAW calculation used
a {\em k\,}-point sampling of \mbox{4$\times$4$\times$6}.  The cluster
method used a O$_{77}$Ti$_{22}$ QM cluster with \mbox{cc-pwCVTZ-DK}
and \mbox{6-311G(2d,p)} basis sets for the inner 3 and outer 19 Ti
atoms, respectively; IGLO-III and IGLO-II basis sets were used for the
inner 15 and outer 52 O atoms, respectively. (Embedding with ePSPs and
point charges was done as described above.)

Calculations for the prototypical perovskites in Table
\ref{tab:prototypical} used the experimental structures as described
in \Ref{ref:pechkis1}.  PBC-GIPAW used a {\em k\,}-point sampling of
6$\times$6$\times$6.  Embedded clusters were constructed as above
(Section \ref{sec:cluster}), and results are given for 21 and 65 QM
atom clusters. The 21 QM atom cluster results are from
non-relativistic calculations using the PW91 DFT functional. The 65 QM
atom clusters are from scalar relativistic PBE calculations.
Differences between PW91 and PBE (not shown in the Table) and
relativistic effects are small, as expected. Tests on the 65 QM atom
ST cluster show that non-relativistic PBE produced $\sigma_{x,y}$ and
$\sigma_{z}$ values that are $\simeq$ 6 and 1.5 ppm more positive,
respectively, than PW91. Adding scalar-relativistic effects changed
$\sigma_{x,y}$ and $\sigma_{z}$ by $\simeq$ +12 and +3 ppm,
respectively, independent of the GGA functional. These effects are
negligible for the corresponding chemical shifts, as expected, due to
cancellation of errors.

The ST and PT 65 QM atom relativistic PBE $\sigma_{x,y}$ and
$\sigma_{z}$ are in good agreement with PBC values, differing at most
by $\simeq$ 33 and 20 ppm respectively.  Isotropic values for both
systems are within 24 ppm of PBC-GIPAW.  A larger discrepancy is seen
in BT, where the cluster values are rigidly shifted by about +50 ppm.
Size effects between the 21 and 65 QM atom results are evident in the
table.  The 21 QM atom values are more shielded [{\em i.e.}, more
positive (see Eq.~(\ref{eq:B_ind}))] than either the 65 QM atom
values or the PBC methods.  Rigidly shifting all 21 QM atom brings
these into better agreement, indicating good cancellation of
errors. This is evident in the chemical shifts shown in
Table~\ref{tab:O-shifts_prototypical} in the next section, where 21
and 65 QM atom clusters are seen to give nearly identical chemicals
shifts.

These results demonstrate that the embedded cluster and PBC-GIPAW
approaches produce comparable agreement with measured isotropic
chemical shifts, regardless of cluster size and methodology.
Differences due to cluster size, DFT functionals, relativistic
effects, and PSPs largely cancel in the chemical shifts, {\em i.e.},
they are absorbed in the constant chemical shielding reference value
$\sigma_{\mathrm{ref}}^{\mathrm {th.}}$.

\begin{table*}[ht]
\begin{center}
\caption{
Calculated oxygen isotropic, axial and anisotropic components (ppm) of the 
chemical shift tensor 
for three PZT 50/50 structural models from  \Ref{ref:mao014105}. 
The notation \mbox{B{-}{-}O{-}{-}B} indicates O atoms with 
two equidistant nn B atoms, and \mbox{B{-}O{-}{-}B} indicates an O atom 
with one short and one long nn B bond.
For cluster-GGA results, numbers in parenthesis show the difference
with GIPAW. For cluster-B3LYP results, numbers in
square brackets show the difference with cluster-GGA values. 
For cases where $\delta_{\rm aniso}=0$ by symmetry, this is indicated by a dash.
}
\begin{tabular*}{0.98\textwidth}{@{\extracolsep{\fill}}lccccccccccc} \hline \hline 
~~~ &\\
              & \multicolumn{3}{c}{$\delta_{\rm iso}$} && \multicolumn{3}{c}{$\delta_{\rm ax}$} &&  \multicolumn{3}{c}{$\delta_{\rm aniso}$} \\
&  \multicolumn{2}{c}{Cluster}       &  GIPAW  &&  \multicolumn{2}{c}{Cluster}       &  GIPAW  &&  \multicolumn{2}{c}{Cluster}      &  GIPAW  \\
                                                         &    B3LYP    &     GGA      &  GGA  &&    B3LYP     &     GGA     &  GGA  &&   B3LYP    &    GGA     &  GGA \\
\hline \\   
ST (cubic) &491 [-3]&494 (-2)&496&&-144 [-11]&-133 (12)&-145&& -- &--& -- \\
\hline \\   
BT (cubic)&536 [2]&534 (-41)&575&&-170 [-16]&-154 (12)&-166&& -- & -- & -- \\
\hline \\   
\multicolumn{12}{l}{BT (P4mm)}    \\[1pt] 
Oax (Ti-O{-}{-}Ti) &591 [20] &571 (-8) &579 &&-207 [-24] &-183 (3) & -186 && --  & -- & -- \\
Oeq (Ti{-}{-}O{-}{-}Ti) &515 [-1] &516 (-19) & 535&&-155 [-14]&-140 (7)&-147&&-24 [2]&-26 (15)&-41 \\
\hline \\  
\multicolumn{12}{l}{PT (P4mm)}    \\[1pt] 
Oax (Ti-O{-}{-}Ti)&644 [35]&608 (2)&606&&-257 [-29]&-228 (8)&-236&& -- & -- & -- \\
Oeq (Ti--O{-}{-}Ti)&449 [-8]&457 (10)&447&&-85 [-10]&-75 (31)&-106&&-22 [7]&-29 (28)&-57 \\
\hline \\   
\multicolumn{12}{l}{PZ(Pbam)}    \\[1pt] 
O1-4g (Z{-}{-}O{-}{-}Z)&357 [-10]&367 (12)&355&&-75 [-5]&-70 (20)&-90&&-33 [1]&-34 (12)&-46 \\
O1'-4g (Z{-}{-}O{-}{-}Z)&346 [-10]&356 (20)&336&&-65 [-4]&-61 (21)&-81&&-14 [2]&-16 (11)&-27 \\
O2-8i (Z-O{-}{-}Z)&355 [-9]&364 (16)&349&&-81 [-4]&-77 (14)&-91&&-4 [1]&-5 (2)&-7 \\
O3-4f (Z{-}{-}O{-}{-}Z)&324 [-11]&336 (26)&309&&-47 [-4]&-43 (9)&-52&&-15 [-1]&-14 (3)&-17 \\
O4-4e (Z{-}{-}O{-}{-}Z)&392 [-6]&399 (-16)&415&&-137 [-7]&-130 (36)&-166&&-14 [1]&-15 (11)&-26 \\
\hline \\    
\multicolumn{12}{l}{PZT (P4mm)}    \\[1pt]       
O1(Zr{-}{-}O{-}{-}Zr) 	                   &	352	[-9]    &	361	(33)	&	328	&&	-72	[-6]	&	-66	(20)	&	-85	&&	-16	[-4]	&	-12	(-4)	&	-8	\\
O2(Zr-O{-}{-}Ti)      	&	398	[7]	&	391\footnotemark[1] (-19)	&	410	&&	-126	[-8]	&	-118	(17)	&	-135	&& --	&	--	&	--	\\
O3(Ti{-}{-}O{-}{-}Ti)   &	418	[-18]	&	436	(9)	&	427	&&	-76	[-11]	&	-65	(21)	&	-86	&&	-59	[4]	&	-63	(38)	&	-101	\\
O4 (Ti-O{-}{-}Zr)   	                   &	679	[40]	&	639	(13)	&	626	&&	-266	[-34]	&	-232	(14)	&	-246	&&--	&-- & --	\\
\hline \\                                    
\multicolumn{12}{l}{PZT (P2mm)}   \\[1pt]                                    
O1(Zr{-}{-}O{-}{-}Zr)	                  &	358	[-9]	&	367	(19)	&	348	&&	-80	[-6]	&	-74	(22)	&	-96	&&	-36	[2]	&	-38	(20)	&	-59	\\
O2(Zr-O{-}{-}Zr) 	                  &	401	[-4]	&	405	(-1)	&	407	&&	-124	[-6]	&	-118	(18)	&	-137	&&	-34	[2]	&	-36	(9)	&	-44	\\
O3(Ti{-}{-}O{-}{-}Ti)	                  &	408	[-11]	&	419	(24)	&	395	&&	-65	[-8]	&	-57	(17)	&	-74	&&	-6	[5]	&	-1	(26)	&	-27	\\
O4(Ti-O{-}{-}Ti)  	                  &	668	[39]	&	629	(12)	&	617	&&	-264	[-32]	&	-232	(6)	&	-238	&&	-49	[3]	&	-46	(8)	&	-54	\\
O5(Ti-O{-}{-}Zr)	                           &	400	[-6]	&	406	(21)	&	385	&&	-81	[-9]	&	-72	(20)	&	-92	&&	-29	[-3]	&	-32	(22)	&	-55	\\
\hline \\                                    
\multicolumn{12}{l}{PZT (R3m)}   \\[1pt]                                    
O1(Zr-O{-}{-}Ti)  	                 &	394	[-7]	&	401	(0)	&	401	&&	-90	[-6]	&	-84	(23)	&	-107	&&	-3	[1]	&	-2      (1)	&	-3	\\
O2(Ti-O{-}{-}Zr)	                          &	466	[-1]	&	467	(11)	&	456	&&	-127	[-13]	&	-114	(21)	&	-135	&&	-2	[1]	&	-1	(-1)	&	-1	\\
\hline \hline
\end{tabular*}
\footnotetext[1]{This cluster-GGA calculation used a smaller Pb cc-pVDZ basis,
  rather than  cc-pVTZ, due to convergence difficulty with the larger basis.}
\label{tab:PZT-delta_components}
\end{center}
\end{table*}

\section{Results}
\label{Results}

In this section, we first describe the calculation of the theoretical
oxygen chemical shielding reference
$\sigma_{\mathrm{ref}}^{\mathrm{th.}}$
[Eq.~(\ref{eq:delta_theory})]. As mentioned, this is done using linear regression of the
calculated chemical shieldings with the corresponding measured
chemical shifts, where available. \cite{ref:Charpentier,ref:Middlemiss2010} 
We next present calculated $^{17}$O chemical shifts for
two perovskite-based B-site alloys, PZT and PMN, using the derived
values of $\sigma_{\mathrm{ref}}^{\mathrm{th.}}$.

\subsection{Determination of the theoretical $^{17}$O chemical shielding reference}

A linear regression was separately evaluated for the PBC and embedded
cluster calculations. \cite{ref:Charpentier,ref:Middlemiss2010}
Additionally for the clusters, separate regressions were performed for
different cluster sizes and DFT functionals.
Table~\ref{tab:O-shifts_prototypical} shows derived
$\sigma_{\mathrm{ref}}^{\mathrm{th.}}$ values for each case, along
with the rms error and maximum deviation in the isotropic chemical
shifts.  The table compares the resulting calculated isotropic
chemical shifts to the measured values.  As in \Ref{ref:pechkis1},
experimental structures were used for all systems except PZ. For PZ,
experimental lattice parameters from neutron scattering measurements
were used together with internal coordinates determined from
first-principles calculations. \cite{ref:Johannes}

Calculated cluster results (21 an 65 QM atoms) with B3LYP and PW91
exchange-correlation are shown together with GIPAW-PBC with PBE
exchange-correlation. The last column shows other GIPAW results, where
available.  For the 65 QM clusters,
$\sigma_{\mathrm{ref}}^{\mathrm{th.}}$ is derived from a more limited
set of calculations, as shown in the Table. For comparison, the effect
of this 
to be only a few ppm.  B3LYP results are seen to give slightly better
agreement with experiment.  Using the same exchange-correlation
treatment, both the 21 and 65 atom QM clusters are seen to give nearly
identical chemicals shifts.  The values of
$\sigma_{\mathrm{ref}}^{\mathrm{th.}}$ in
Table~\ref{tab:O-shifts_prototypical} are used below to determine the
theoretical chemical shifts for PZT and PMN in Tables
\ref{tab:PZT-delta_components} and \ref{tabPMN-delta_components}.
Given the small difference between the PW91 and PBE GGA functionals
and cancellation effects in chemical shifts, both PW91 and PBE are
labeled as GGA in all further results below.

\subsection{Results for Pb(Zr$_{{1-x}}$Ti$_{x}$)O$_3$ (PZT) and Pb(Mg$_{1/3}$Nb$_{2/3}$)O$_3$ (PMN)}

PZT is a homovalent mixture of Ti$^{4+}$ and Zr$^{4+}$ transition
metal cations, while Pb(Mg$_{1/3}$Nb$_{2/3}$)O$_3$ (PMN) is a
heterovalent 2:1 mixture of the Nb$^{5+}$ transition metal cation and
the Mg$^{2+}$ alkaline earth cation. Both embedded cluster and GIPAW
calculations were carried out for PZT, while PMN results were obtained
only with the GIPAW method, using PBC.

Disordered PZT 50/50 was modeled using first-principles relaxed
structures \cite{ref:mao014105} with different imposed B-site ordering
and symmetry: i) [001] ordering parallel to the ferroelectric
polarization, with tetragonal P4mm symmetry
(\mbox{a\,$\times$\,a\,$\times$\,2c; c/a = 1.045}); ii) [001] ordering
perpendicular to the [100] ferroelectric polarization, with
orthorhombic P2mm symmetry
(\mbox{a$^\prime$\,$\times$\,a\,$\times$\,2a; a$^\prime$/a = 1.04});
iii) [111] ordering (rocksalt B sublattice) parallel to the
ferroelectric polarization, with R3m symmetry.  Theoretical PZT
isotropic, axial, and anisotropic chemical shifts are summarized in
Table~\ref{tab:PZT-delta_components}.  [As mentioned, the axial and
anisotropic components in Eq.~(\ref{eq:sigmaiso}) were calculated
using the convention that the ``$z$'' principal axis is that most
nearly parallel to the \mbox{B-O-B} bond direction of the target O
atom.]  For comparison, results for ST, BT, PT and PZ are reproduced
from Table~\ref{tab:O-shifts_prototypical}.  For each inequivalent
target O atom in the above structural models, the corresponding 21 QM
atom embedded cluster was used to calculate the chemical shielding
tensor with both B3LYP and GGA exchange-correlation.  Results from
PBC-GIPAW with GGA exchange-correlation are also shown.  Chemical
shifts were determined using the corresponding values of
$\sigma_{\mathrm{ref}}^{\mathrm{th.}}$ in
Table~\ref{tab:O-shifts_prototypical}.  The GGA 21 QM atom embedded
cluster and PBC-GIPAW RMS errors differ by 4 ppm.  As seen in
Tables~\ref{tab:O-shifts_prototypical}, the 21 QM atom embedded
cluster chemical shifts are also in excellent agreement with those
from the larger 65 QM atom cluster.

PMN PBC-GIPAW calculations were based on a low symmetry
(\mbox{3\,$\times$\,2\,$\times$\,2}) 60-atom perovskite supercell
structure with relaxed internal coordinates, \cite{ref:PMN} with
B-site cations arranged according to the random-site
model. \cite{randomsite} X-ray patterns of well-annealed PMN samples
indicate a homogeneous average structure, which exhibits rocksalt-like
1:1 B-site ordering, which is well described by the random-site model.
Locally, the random-site model corresponds to B-site [111] planes,
alternating between pure Nb and mixed Nb/Mg layers.  Thus there are
twice as many \mbox{Nb-O-Mg} as \mbox{Nb-O-Nb} O-sites in the 60-atom
supercell, since the Mg atoms all reside in the mixed [111] planes.
Similarly, there are no \mbox{Mg-O-Mg} O-sites.  Unlike earlier
models, the random-site model satisfies charge neutrality locally. The
previously accepted space-charge model was based on the apparent
inability to fully anneal samples.  \cite{randomsite} The 60-atom
perovskite supercell structure, \cite{ref:PMN} used in the present
calculations, is consistent with the random-site model.  Grinberg {\it
  et al.} \cite{ref:PMN} found good agreement between this relaxed
60-atom supercell structure and pair distribution functions (PDFs)
obtained by neutron scattering experiments. \cite{PMN_PDF_1998} This
indicates that this structural model reasonably represents the local
structure in PMN.  Calculations for PMN were done only with PBC-GIPAW,
and the results are shown in Table~\ref{tabPMN-delta_components}.

\section{Discussion}
\label{Discussion}

As previously noted in \Ref{ref:pechkis1}, there is a large anisotropy
between the $\sigma_{x,y}$ and $\sigma_z$ principal values in
Table~\ref{tab:prototypical}.  The $\sigma_{x,y}$ principal values are
large and negative (deshielded), while $\sigma_z$ values tend to be
considerably smaller and positive (shielded).  (As mentioned, the
``$z$'' principal axis is identified as that most closely aligned with
the B-O-B bond direction of the target O atom.)  The present
calculations show that this anisotropy is also found in PZT and PMN
and is reflected in the large $\delta_{ax}$ values
(Eq.~\ref{eq:sigmaiso}) in Tables~\ref{tab:PZT-delta_components} and
\ref{tabPMN-delta_components}.

As shown in \Ref{ref:pechkis1} for prototypical perovskites, the large
$\delta_{ax}$ values are due to paramagnetic contributions to
$\sigma_{x,y}$ from virtual transitions between O(2p) and unoccupied
B($n$d) states. The p-d hybridization contributes predominantly to
$\sigma_{x,y}$, due to O atoms having only two nearest neighbors in
perovskites, with linearly arranged \mbox{B-O-B} structural units.  As
the \mbox{B-O-B} bond distances vary, large variations in the chemical
shielding can occur, resulting in a strong dependence on
$r_\mathrm{s}$.  We find a similar dependence for the PZT and PMN
alloy systems.

\subsection{PZT}

\begin{figure}[ht]
\begin{center}
\includegraphics[scale=0.30]{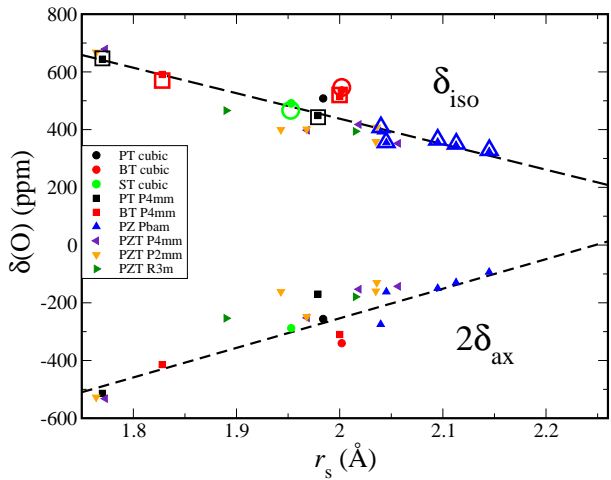}
\caption{(Color online) Calculated PZT oxygen isotropic
  $\delta_\mathrm{iso}$ and axial $2\delta_\mathrm{ax}$ chemical
  shifts (B3LYP 21 QM atom cluster values from
  Tables~\ref{tab:PZT-delta_components}), plotted as a a function of
  $r_\mathrm{s}$, the shortest B-O bond length of the targeted O atom.
  For comparison, calculated results for ST, BT, PT, and PZ and from
  experiment (hollow symbols) are also shown. \cite{ref:pechkis1} The
  dashed straight lines are linear fits to calculated values for ST, BT,
  PT, and PZ from \Ref{ref:pechkis1}.  }
\label{fig:CS_vs_BL}
\end{center}
\end{figure}

Figure~\ref{fig:CS_vs_BL} plots PZT
 isotropic and axial
chemical shifts as a function
of $r_\mathrm{s}$, the shortest B-O bond length of the targeted O
atom.  
The B3LYP 21 QM atom cluster results from Table~\ref{tab:PZT-delta_components} 
are plotted.
For comparison, results for 
ST, BT, PT, and PZ
and from experiment are also shown. \cite{ref:pechkis1} 
The dashed straight lines are the linear fits
to the calculated results for these prototypical perovskites, taken from \Ref{ref:pechkis1}.
A plot of PBC-GIPAW results from Table~\ref{tab:PZT-delta_components}
(not shown) is very similar, which is consistent with the generally good
agreement of the chemical shift results in the Table
between the two calculational
approaches.
A nearly linear dependence on $r_\mathrm{s}$ (with slope $\sim 850$ ppm/{\AA}) is seen in both
$\delta_\mathrm{iso}$  and $\delta_\mathrm{ax}$, across all the
systems studied.
The axial shift is plotted as 
\mbox{$2\delta_\mathrm{ax} = \delta_z - \delta _{{\rm{iso}}}$} 
(Eq.~\ref{eq:sigmaiso}) to emphasize that the 
linear dependence is largely due to $\delta_{x,y}$, while
$\delta_{z}$ has a much weaker dependence on  $r_\mathrm{s}$,
as previously noted for the prototypical perovskites.  \cite{ref:pechkis1} 
As seen in the figure, the calculated PZT 50/50 results follow the
same trends as in  \Ref{ref:pechkis1}. 

Pb(Zr$_{{1-x}}$Ti$_{x}$)O$_3$ NMR $^{17}$O magic angle spinning  (MAS)  
central peak spectra 
were presented by Baldwin {\em et al.}
\cite{ref:Baldwin05} for a range of concentrations $x$.
Tabulated chemical shifts were given 
only for the endpoint PT and PZ
compounds (reproduced here in Table~\ref{tab:O-shifts_prototypical}), 
whose spectra consist of well-defined narrow peaks.
MAS
removes broadening due to chemical shift anisotropy in powder
samples, but only partly averages second-order quadrupolar broadening.
The narrow peaks in the endpoint PT and PZ compounds indicate that electric field gradients (EFGs) at the  $^{17}$O nuclei
are small. \cite{ref:Baldwin05} This is consistent with first-principles calculations of O EFGs. 
\cite{ref:Johannes,ref:mao014105} Indeed, the  $^{17}$O peak positions
are within a few ppm of the experimental isotropic chemical
shifts. \cite{ref:Baldwin05} 
There are  two inequivalent O atoms in PT, two ``equatorial'' O$_{\rm
  eq}$ (which has two equidistant
nn Ti atom) and one ``axial'' O$_{\rm ax}$ atom (which has one short and one long Ti-O
bond).
A narrow peak at $443$\,ppm has twice the (integrated) relative intensity of the peak
at $647$\,ppm, and these were assigned to the  O$_{\rm eq}$ 
and  O$_{\rm ax}$ atoms, respectively. These are in good agreement
with the calculated results in Table~\ref{tab:O-shifts_prototypical}.
The B3LYP calculation accurately reproduces this splitting,
while GGA underestimates it by $\simeq 45$\,ppm.
In PZ, there are five inequivalent O sites, 
which corresponds to five MAS peaks,  centered near 350\,ppm, within $\pm \sim 40$\,ppm.
Measured and
calculated values in Table~\ref{tab:O-shifts_prototypical} are in good agreement.

\begin{table}[ht]
\caption{
GIPAW calculated oxygen isotropic, axial and anisotropic components (ppm) of the 
chemical shift tensor 
for PMN. 
A 60-atom supercell with relaxed internal coordinates was used. \cite{ref:PMN} 
The \mbox{B{-}O{-}{-}B$^{\prime}$} bond notation is the same as in Table~\ref{tab:PZT-delta_components};
bond lengths are categorized as equidistant if they differ by less than 0.05 {\AA}.}
\begin{tabular*}{0.48\textwidth}{@{\extracolsep{\fill}}lccccc} \hline \hline\\
           & Nb{-}O  & B$^{\prime}${-}O  &   $\delta_{\rm iso}$ & $\delta_{\rm ax}$ & $\delta_{\rm aniso}$  \\
           &     \AA  & \AA        &                   &                  &                     \\\hline
           &          &            &                                       &                  &                     \\
Nb{-}O{-}Mg   & & & & & \\
           &	2.11  & 2.09	   &             235      &      -6	   &              -5     \\
           &    2.11  & 2.09       &             235      &      -7          &              -2      \\
           &	2.06	&	2.08	&		280	&	-26	&	-47	\\
           &	2.07	&	2.09	&		277	&	-22	&	-43	\\\hline
           &          &            &                             &                  &                      \\
Nb{-}O{-}{-}Mg   & & & &  & \\
&	1.89	&	2.06	&		419	&	-109	&	-27	\\
&	1.90	&	2.07	&		405	&	-97	&	-11	\\
&	1.90	&	2.09	&		390	&	-110	&	-25	\\
&	1.90	&	2.11	&		387	&	-110	&	-31	\\
&	1.91	&	2.11	&		390	&	-111	&	-83	\\
&	1.93	&	2.08	&		362	&	-91	&	-7	\\
&	1.93	&	2.09	&		368	&	-94	&	-10	\\
&	1.93	&	2.10	&		374	&	-100	&	-83	\\
&	1.94	&	2.13	&		369	&	-75	&	-36	\\
&	1.94	&	2.18	&		390	&	-75	&	-17	\\
&	1.95	&	2.11	&		362	&	-81	&	-11	\\
&	1.95	&	2.18	&		385	&	-67	&	-18	\\
&	1.96	&	2.05	&		346	&	-78	&	-29	\\
&	1.96	&	2.12	&		348	&	-61	&	-44	\\
&	1.97	&	2.05	&		338	&	-71	&	-36	\\
&	1.97	&	2.10	&		343	&	-64	&	-11	\\
&	1.97	&	2.13	&		327	&	-53	&	-13	\\
&	1.98	&	2.06	&		340	&	-51	&	-47	\\
&	1.99	&	2.13	&		316	&	-43	&	-7	\\
&	2.01	&	2.06	&		310	&	-36	&	-39	\\\hline
        	&		&			&		&		&		\\
Nb{-}O{-}{-}Nb  & &  & & & \\
&	1.86	&	2.23	&		436	&	-130	&	-21	\\
&	1.87	&	2.14	&		421	&	-130	&	-3	\\
&	1.87	&	2.14	&		421	&	-129	&	-8	\\
&	1.87	&	2.23	&		437	&	-130	&	-18	\\
&	1.89	&	2.20	&		416	&	-114	&	-27	\\
&	1.90	&	2.20	&		414	&	-110	&	-24	\\
&	1.91	&	2.11	&		394	&	-111	&	-23	\\
&	1.91	&	2.11	&		392	&	-108	&	-18	\\
&	1.94	&	2.03	&		401	&	-122	&	-8	\\
&	1.95	&	2.01	&		399	&	-122	&	-11	\\ \hline 
        	&		&			&		&		&		\\
Nb{-}O{-}Nb  & &  & & & \\
&	2.06	&	2.09	&		347	&	-62	&	-9	\\
&	2.08	&	2.08	&		340	&	-56	&	-11	\\\hline\hline
\end{tabular*}
\label{tabPMN-delta_components}
\end{table}

\vspace{11pt}

At intermediate Ti and Zr concentrations, the spectra in Fig. 3 of
\Ref{ref:Baldwin05} show that only a few of the narrow endpoint peaks
persist.  As Zr is added to PT, the narrow $647$~ppm PT peak decreases
quickly in intensity. It can no longer be observed in the $x=0.55$
sample.  A broad feature, between about 350 and 450 ppm is fully
developed near $x \simeq 0.50$, with narrower embedded features at
$\simeq$~370 and 430\,ppm.  This broad feature distribution of
inequivalent O-sites in the disordered PZT solid solution samples.
The 430~ppm feature, which is close to the PT O$_\mathrm{eq}$ 447~ppm
peak, is observed to persist down to 25\% Ti concentration. Baldwin
{\it et al.}~\cite{ref:Baldwin05} assign the 430~ppm feature to a site
similar to that of the PT O$_\mathrm{eq}$ atom, {\it i.e.} a locally
Ti-O$_\mathrm{eq}$-Ti (undimerized) chain-like configuration.  They
further suggest that this peak could indicate Ti clustering on a
spatial scale of at least two unit cells in PZT.

Our results suggest an alternative explanation for the persistence of
the observed 430~ppm feature.  Our calculations show similar chemical
shifts for a range of B-O-B$^{\prime}$ environments, with
$r_\mathrm{s}$ ranging between $\sim 2.0 - 2.1$ {\AA} in all the PZT
50/50 structural models, as seen in
Table~\ref{tab:PZT-delta_components} and Fig.~\ref{fig:CS_vs_BL}.
Thus Ti clustering need not be invoked to explain the persistence of
the 430~ppm feature in the measured spectra.

The apparent disappearance, at intermediate concentrations, of the
647\,ppm peak does, however, indicate a reduced occurrence of a
PT-like O$_{\rm ax}$ site with a short 1.77 {\AA} Ti-O bond.  We find
such a site only in the relaxed P4mm and P2mm PZT 50/50 simulations.
The P4mm and P2mm models have [100] type B-site ordering.  The absence
of the 647\,ppm peak in the measurements indicates that local
occurrences of [100] type B-site ordering are rare.  Instead, the lack
of the 647\,ppm peak in our R3m PZT 50/50 simulations suggests that
local rocksalt-like B-site ordering is more prevalent in disordered
PZT.  This conclusion is also supported by the R3m structural model
having the lowest total energy of all the structural models by $\sim$
23 mRy. \cite{ref:mao014105}  Moreover, the calculated Ti EFG's for
50/50 R3m were significantly smaller and in better agreement with
measured values, than the other B-site orderings. \cite{ref:mao014105}

Baldwin {\em et al.}  \cite{ref:Baldwin05} also remark a narrow
287\,ppm peak that appears in their $x=0.48$ sample. This peak is
broader in $x=0.55$ and $x=0.25$ samples and is absent in PZ. These
authors note that $x=0.48$ is the composition corresponding to the
morphotropic phase boundary (MPB) and conjecture that the 287 ppm peak
evidences a new oxygen environment in a distinct crystalline
monoclinic phase, which has been suggested to bridge the
MPB. \cite{ref:Baldwin05} They suggest that the new environment at
$x=0.48$ is a Ti-O-Zr site, which becomes ordered in the
crystallographic sense as the long-range order of the monoclinic phase
is established.  As indicated by Fig.~\ref{fig:CS_vs_BL}, isotropic
chemicals shifts near 287\,ppm are associated with large $r_{\rm s}
\simeq 2.15$\,{\AA}.  As most Zr-O bond lengths are larger than those
of Ti-O in the PZ and PZT 50/50 structural models, the 287\,ppm
feature could also be attributed to Zr-O-Zr sites with bond lengths
distributed near $\simeq 2.15$\,{\AA}.

\begin{figure}[t]
\begin{center}
\includegraphics[scale=0.30]{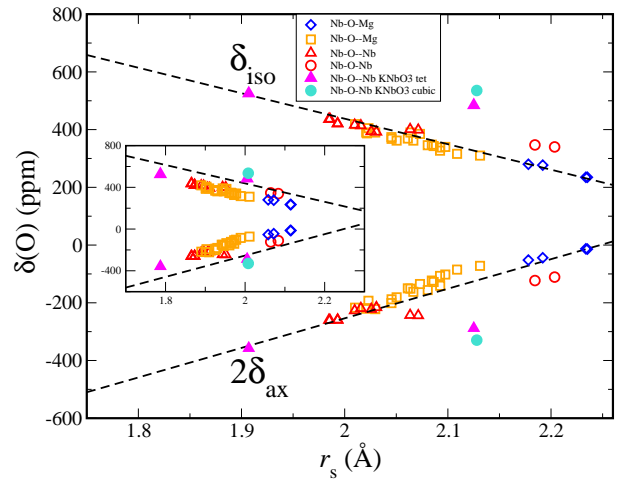}
\caption{(Color online) GIPAW calculated PMN oxygen isotropic
  $\delta_\mathrm{iso}$ and axial $2\delta_\mathrm{ax}$ chemical
  shifts, plotted as a function of $r_\mathrm{s}$.  The PMN chemical
  shifts are from Table~\ref{tabPMN-delta_components}, and the symbols
  indicate the B-O-B$^{\prime}$ configuration, using the convention in
  that table. For comparison, results for cubic and tetragonal
  KNbO$_3$ are also shown.  The dashed lines are the same as that in
  Fig.~\ref{fig:CS_vs_BL}.  As plotted, the $r_\mathrm{s}$ values have
  been increased by 0.12 {\AA} (see text). The unshifted values are
  shown in the inset.}
\label{fig:CS_vs_BL_PMN}
\end{center}
\end{figure}

\subsection{PMN}
PMN, by contrast with PZT, is a heterovalent 2:1 mixture of the
Nb$^{5+}$ transition metal cation and the Mg$^{2+}$ alkaline earth
cation.  Figure~\ref{fig:CS_vs_BL_PMN} plots
Pb(Mg$_{1/3}$Nb$_{2/3}$)O$_3$ (PMN) isotropic and axial chemical
shifts (GIPAW results from Table~\ref{tabPMN-delta_components}) as a
function of $r_\mathrm{s}$, the shortest B-O bond length of the
targeted O atom. As plotted, the PMN $r_\mathrm{s}$ values, have been
increased by \mbox{0.12 {\AA}} for both $\delta_\mathrm{iso}$ and
$\delta_\mathrm{ax}$, as further discussed below.  The unshifted
values are plotted in the inset.  The dashed lines are the same as
that in Fig.~\ref{fig:CS_vs_BL}.  With the \mbox{0.12 {\AA}} rigid
shift for all O-sites, the PMN chemical shifts are seen to follow the
same linear trend as in the homovalent B-site systems in
Fig.~\ref{fig:CS_vs_BL}.  For \mbox{Nb-O-Mg} coordinated O-atoms,
$r_\mathrm{s}$ is taken as the Nb-O bond length, since the
hybridization mechanism does not apply to Mg, which has no low-lying
unoccupied d-states. (As mentioned, the linear dependence is due to
paramagnetic contributions to $\delta_\mathrm{xy}$ from virtual
transitions between O(2p) and unoccupied B($n$d) states.) We are not
aware of published NMR $^{17}$O spectra for PMN.

Although the linear dependence of $\delta_\mathrm{iso}$ and
$\delta_\mathrm{ax}$ in Fig.~\ref{fig:CS_vs_BL_PMN} have the same
slope as the homovalent B-site systems, a 0.12 {\AA} $r_\mathrm{s}$
offset is needed for the PMN values to fall on the same line.  As
mentioned, the linear dependence reflects variations in the magnitude
of the paramagnetic O(2p)-Nb(4d) hybridization contributions to the
$\delta_{x,y}$ principal values.  This indicates that the effective
$r_\mathrm{s}$ is controlled by the spatial extent of the paramagnetic
screening currents.  The Nb$^{5+}$ cation could be expected to modify
the spatial extent of these currents, compared to the B$^{4+}$ cations
in the homovalent systems, due to the larger electrostatic attraction
of the Nb$^{5+}$ cation.  The 0.12 {\AA} offset in PMN renormalizes,
in effect, the strength of the O(2p)-Nb(4d) hybridization.  This
observation would appear to indicate that for these systems the
$r_\mathrm{s}$-dependence of $\delta_{x,y}$ is given by
$\delta_{x,y}=m(r_\mathrm{s}-r_0)$, where the slope $m$ is nearly the
same for all transition-metal coordinated O atoms in perovskites,
while the intercept $r_0$ depends on other factors, such as the ionic
charge of the nearest neighbor cation.  To further examine this, we
performed GIPAW calculations for cubic and tetragonal KNbO$_3$,
\cite{ref:KNbO3structure} which are also plotted in
Fig.~\ref{fig:CS_vs_BL_PMN}.  While the tetragonal axial O values
(short Nb-O $r_\mathrm{s}$) are consistent with the linear trend, the
KNbO$_3$ tetragonal equatorial and cubic O results (two equidistant
Nb-O bonds) show significant deviations.  [Similar deviations occur in
  Fig.~\ref{fig:CS_vs_BL} for the tetragonal equatorial and cubic O BT
  results. We note that both K and Ba have larger crystalline ionic
  Shannon\cite{Shannon1976} radii, 1.78 and 1.75 {\AA}, respectively,
  than Sr or Pb, 1.58 and 1.63 {\AA}, respectively.]  Smaller
deviations are also seen in the two largest $r_\mathrm{s}$ values for
the PMN oxygen atoms with two nearly equidistant Nb atoms, but for
somewhat larger values of $r_\mathrm{s}$ than in KNbO$_3$.  In the
case of heterovalent B-site perovskites, the covalency of A-site atoms
could also be important, as indicated by the smaller deviations, for
long $r_\mathrm{s}$, in PMN compared to KNbO$_3$. These observations
warrant further investigation to clarify these issues.

PMN is an end-point of the solid-solution series
\mbox{$(x)$PbTiO$_3$$-$$(1-x)$Pb(Mg$_{1/3}$Nb$_{2/3}$)O$_3$} (PMN-PT).
In the 60-atom PMN structural model, there are no Nb-O bonds as small
as the short Ti-O$_{\rm ax}$ bond $\simeq 1.7$\,{\AA} in PT, which is
associated with its high degree of tetragonality ($c/a=1.065$). This
short bond corresponds to the large $\delta_\mathrm{iso} \simeq
640$\,ppm, which is also seen in the P4mm and P2mm PZT structural
models, both of which also show a high degree of tetragonality.  The
tetragonality of PMN-PT decreases as the Ti concentration is reduced
from PT-rich compositions, and the average symmetry switches from
tetragonal to rhombohedral at the morphotropic phase boundary (MPB)
$x\simeq 0.35$.  The largest piezoelectric response is typically
achieved at concentrations near the MPB.~\cite{ref:Pa97} Polarization
rotation has been proposed as the origin of the large piezoelectric
response at the MPB, via intermediate monoclinic phases,
\cite{ref:PZT-PhaseTrans,ref:FC00,ref:Wu-Krakauer-03} where the
tetragonality increases as the polarization rotates from [111] to
[100] (pseudocubic) directions, with applied electric field along a
pseudocubic axis.  Increased tetragonality, compared to PT, has been
seen in some other perovskite based solid solutions, such as some Bi
based materials.~\cite{tetragonality-PT-Bi-2005,tetragonality-PT-Bi}
The present calculations indicate that $^{17}$O NMR chemical shift
measurements could be a useful probe in this regard, as increased
tetragonality is accompanied by shortened transition-metal/oxygen
bonds.

\section{Summary}
\label{summary}

First-principles oxygen NMR chemical shift tensors were calculated for
PZT and PMN, which are representative, respectively, of homovalent and
heterovalent perovskite-structure B-site alloys.  Quantum chemistry
methods for embedded clusters and the GIPAW method for periodic
boundary conditions were used.  Results from both methods are in good
agreement for PZT and prototypical perovskites.  PMN results were
obtained using only GIPAW.  Both isotropic and axial chemical shift
components were found to vary approximately linearly as a function of
the nearest-distance transition-metal/oxygen bond length,
$r_\mathrm{s}$.  Using these results, we argue against Ti clustering
in PZT, as conjectured from recent $^{17}$O NMR measurements.  Our
findings indicate that $^{17}$O NMR measurements, coupled with
first-principles calculations, can be an important probe of local
structure in complex perovskite solid solutions.

\section{Acknowledgments}
This research was supported by Office of Naval Research grants
N00014-08-1-1235 and N00014-09-1-0300.  DLP acknowledges partial
support from a Virginia Space Grant Consortium Graduate Research
Fellowship.  Support for computations was provided in part by National
Science Foundation TeraGrid resources at the National Center for
Supercomputing Applications (NCSA) under grant number TG-DMR100024.
Additional computational resources were provided by the Center for
Piezoelectric by Design. We acknowledge useful discussions with Gina
Hoatson and Robert L. Vold.

\linespread{1}

\bibliography{paper} 
\bibliographystyle{ieeetr} 

\end{document}